\documentclass[prb,twocolumn]{revtex4}
\usepackage{caption}
\usepackage{graphicx}
\usepackage{subfigure}
\usepackage{epsfig}
\usepackage{amsmath}
\usepackage{amsfonts}
\usepackage{amssymb}
\usepackage[version=3]{mhchem}
\providecommand{\U}[1]{\protect\rule{.1in}{.1in}}

\begin{document}

\title{Field-Induced Weyl-Like Crossings and Longitudinal Magnetoconductivity in Dirac Systems}

\author{J. L. Acosta Avalo}

\affiliation{%
\textit{Instituto Superior de Tecnolog\'{\i}as y Ciencias Aplicadas, Universidad de La Habana, Ave. Salvador Allende No. 1110, Vedado, La Habana 10400 Cuba}\\
\textit{Instituto de Cibern\'{e}tica, Matem\'{a}tica y F\'{\i}sica, Calle E esq. 15, No. 309, Vedado, La Habana 10400 Cuba}
}

\begin{abstract}

\small{We investigate the field-induced separation of effective Weyl-like crossings and the associated longitudinal magnetoconductivity in Dirac systems using a finite-temperature quantum electrodynamics  response-function approach. Starting from the one-loop photon polarization tensor in a magnetized fermionic medium, we identify a longitudinal electromagnetic mode associated with a nonvanishing pseudoscalar invariant \(\mathbf E\cdot\mathbf B\). We use this response to construct an effective low-energy Dirac model in which the two chiral sectors undergo opposite momentum shifts. Within this description, we derive an analytical expression for the longitudinal magnetoconductivity without introducing an independent phenomenological internode relaxation time. The chiral contribution contains an explicit \(B^2\) prefactor, while its coefficient depends on the magnetic field and temperature through Landau-level thresholds and Fermi–Dirac occupation factors. The resulting conductivity exhibits a nonmonotonic magnetic-field dependence and an inverse-temperature regime under appropriate conditions. These results provide a connection, within the proposed effective model,  between the longitudinal electromagnetic response, field-induced Weyl-like crossings, and longitudinal magnetotransport.
}

\end{abstract}

\maketitle

\section{Introduction}
\label{Intro}

Dirac and Weyl semimetals provide condensed-matter realizations of
relativistic fermionic quasiparticles, whose low-energy band structures give rise to
unconventional electromagnetic and transport
phenomena\cite{Armitage2018,Mele2012,Miransky2013,Miransky2014,Zyuzin2017,Min2024,Wang2012,Wang2013,Borisenko2014,Liu2014,Tanay2020,Kharzeev2016,Burkov2011,Chen2013}. Of particular interest is the chiral
anomaly\cite{Adler1,Bell}, which allows parallel
electric and magnetic fields to generate a
nonequilibrium axial-charge density and an associated
contribution to the longitudinal electrical
conductivity. This
contribution is commonly characterized by a
quadratic magnetic-field dependence\cite{Armitage2018,Spivak2013,Li2015},
$\sigma_B\propto B^2$.
Enhanced longitudinal magnetoconductivity has
been reported in several Dirac and Weyl
semimetals, including Na$_3$Bi and
ZrTe$_5$~\cite{Xiong2015,Kharzeev2016}.

Theoretical approaches based on semiclassical
kinetic theory, Berry-curvature corrections,
and anomalous transport have established a
connection between chiral charge pumping and
longitudinal magnetoconductivity
\cite{Spivak2013,Burkov2015,Goswami2015,Johansson2019}. In conventional steady-state descriptions,
the Weyl-node structure is specified by the
electronic Hamiltonian, whereas the
nonequilibrium axial charge is controlled
by internode relaxation.
\cite{Spivak2013,Armitage2018,Ong2021}. These approaches
describe transport in systems with an
established chiral band structure. A complementary question is whether the longitudinal electromagnetic mode emerging from the response of a magnetized Dirac system can induce an effective separation of Weyl-like crossings in momentum space, and how this field-induced structure manifests itself in the longitudinal magnetoconductivity.

Here we investigate this question within
finite-temperature quantum electrodynamics (QED).
Starting from the one-loop photon polarization
tensor of a magnetized fermionic medium,
we consider the longitudinal electromagnetic
mode propagating parallel to the external
magnetic field. The electric field associated
with this mode induces a nonvanishing pseudoscalar invariant 
$\mathbf{E}\cdot \mathbf{B}$,
which enters the axial-current response.
We take the relativistic QED response as the microscopic starting point for an effective low-energy Dirac model. 
Within this model, an effective axial coupling produces opposite momentum shifts in the two chiral sectors, resulting in a field-dependent separation of the zero-energy crossings of the effective spectrum  rather than to the equilibrium band structure of a specific material. We refer to these crossings as Weyl-like because their topological character is not established independently in the present analysis. The relevant longitudinal mode represents a collective electromagnetic
excitation rather than a freely propagating photon in the material.

Unlike formulations that prescribe a chiral
chemical potential~\cite{Fukushima2008,Kharzeev2016} $\mu_5$, our approach does not require an externally imposed chiral asymmetry. Although we introduce an ordinary electromagnetic chemical potential $\mu$ (as in Ref.~\cite{Vilenkin} and Ref.~\cite{Acosta2016} ), our results remain valid for $\mu=0$. Instead, the chiral imbalance is induced dynamically by the
absorptive part of the electromagnetic
response function, without introducing an
independent internode relaxation time. This approach therefore provides a complementary description of the longitudinal response, rather than a complete account of disorder-dependent steady-state transport. To clarify the nature of the chiral imbalance\cite{Acosta2016}, we distinguish between two types of chiral symmetry breaking:  explicit breaking due to a nonzero mass, which is compatible with thermodynamic equilibrium, and anomalous breaking associated with the pseudoscalar invariant $\mathbf{E} \cdot \mathbf{B}$. The latter is associated with nonequilibrium processes involving electric currents and charge transport. In the present discussion, the chiral imbalance is understood in terms of the momentum-space distributions of particles and antiparticles with a common helicity. Its dynamical generation is associated with the nonvanishing pseudoscalar invariant \(\mathbf{E}\cdot\mathbf{B}\), which acts as a source term in the axial-current continuity equation~\cite{Acosta2016}.

The resulting longitudinal magnetoconductivity
contains an explicit $B^2$ contribution whose
coefficient depends on the magnetic field and
temperature through Landau-level thresholds
and Fermi--Dirac occupation factors.
Consequently, the full conductivity does
not generally obey a quadratic magnetic-field
law. In an appropriate high-temperature
regime, the coefficient exhibits an
inverse-temperature dependence, while
its behavior near Landau-level thresholds
is nonmonotonic. These features provide
possible experimental signatures of the
longitudinal electromagnetic response
considered here.

The remainder of this paper is organized
as follows. Sec. \ref{sec2} summarizes the
longitudinal electromagnetic response
and axial current in a magnetized
relativistic medium. Unless otherwise stated, the QED expressions are written in natural units. In the effective condensed-matter spectrum and numerical estimates, we restore \(\hbar\) and \(c\) and use Gaussian CGS electromagnetic units. Sec. \ref{sec3},
we construct the low-energy Dirac model,
obtain the longitudinal magnetoconductivity,
and discuss its magnetic-field and
temperature dependence. Sec. \ref{sec4}
presents our conclusions.

\section{LONGITUDINAL PHOTONS AND CHIRAL CURRENT IN A MAGNETIZED RELATIVISTIC MEDIUM}
\label{sec2}

We now summarize the essential QED results for a magnetized relativistic plasma at finite temperature \(T\) (expressed in energy units) and chemical potential \(\mu\), as derived in detail in our previous work\cite{Acosta2015,Acosta2016}. These results provide the basis for the effective low-energy description of Dirac systems developed in Sec. \ref{sec3}.

We consider a relativistic electron-positron plasma in a constant magnetic field $\mathbf{B}$ parallel to the $x_3$ axis. For $\mu\neq0$, an ionic positive background ensures charge neutrality, although its dynamical effects can be neglected. The fermion spectrum consists
 of Landau levels:

 \begin{equation}\label{espectro/energia}
   \varepsilon_{n_{l},p_{3},s_3}=\sqrt{p^{2}_{3}+m^{2}+|e|B(2n_{l}+1-sgn(e) s_{3})},
\end{equation}

\noindent where $p_3$ is the momentum along $\mathbf{B}$, $m$ the electron mass, $s_3=\pm1$ the spin eigenvalues along $x_3$, and $n_l=0,1,\ldots$ the Landau index. States are two-fold spin degenerate except the ground state ($n_l=0$), where $s_3=-1$ for electrons and $s_3=+1$ for positrons. In our description, we retain only the lowest Landau levels. Higher levels are neglected when their thermal occupation is exponentially small, \(\exp[-(\varepsilon_{n_{l},p_{3},s_3}\mp\mu)/T]\ll 1\). Equivalently\cite{John,prl2000}, in terms of the Landau scale \(\Delta_B = \sqrt{|e|B}\), this regime corresponds to \(\Delta_B \sim \mu\) and \(\Delta_B \gg T\).

The photon field $a_\mu$ ($E\ll B$) satisfies the Schwinger-Dyson equation in Fourier space (with $\nu=1,2,3,4$)\cite{Fradkin2}:

\begin{equation}\label{ecuacion/SD}
  [k^{2} g_{\mu\nu}-\Pi_{\mu\nu}(k|A_{\mu}^{\text{ext}})]a^{\nu}(k)=0,
\end{equation}

\noindent where $g_{\mu\nu}=(1,1,1,-1)$ is the Minkowski metric, $\Pi_{\mu\nu}(k|A_{\mu}^{ext})$ is the one-loop photon self-energy in a magnetized medium \cite{Shabad1,Hugo2,Hugo3,Shabad}, and  $k^2=k_3^2+k_\perp^2-\omega^2$. Here $k_3$ and $k_\perp$ are the components of the photon
four-momentum parallel and perpendicular to $\mathbf B$, respectively,
and $\omega$ is its energy. In the medium, the photon self-energy $\Pi_{\mu\nu}$ admits three eigenmodes  $b_\mu^{(i)}$ with dispersion law  $k^{2}=\eta_{i}(k_{3},\omega,k_{\bot};B,T,\mu)$, where $\eta_i$ ($i=1,2,3$) are the eigenvalues\cite{Hugo2,Shabad1}, in both the $C$-invariant and non-invariant cases: two transverse\cite{Hugo6,Lidice} and one longitudinal \cite{Hugo3,Acosta2016}
 (see Appendix~\ref{App1} for details). For propagation parallel to $\mathbf{B}$, the longitudinal mode\cite{Acosta2015,Acosta2016} is given by the pseudovector $b_\mu^{(2)}(k)=aR_2 (F^*k)_\mu$, where $R_2=1/Bz_1^{1/2}$ is a normalization factor and $F^{*}_{\mu\nu}$ is the dual of the electromagnetic tensor
$F_{\mu\nu}$. The parameter $a$ has dimension of vector potential and is determined by the applied perturbative electric field parallel\cite{Acosta2015,Acosta2016} to $\mathbf{B}$:

\begin{equation}\label{campo electrico pseudovectorial}
\mathbf{E}_B = E^{(2)}\mathbf{e}_B = a (k_3^2-\omega^2)^{1/2}\mathbf{e}_B,
\end{equation}

\noindent where $E^{(2)}=a(k_3^2-\omega^2)^{1/2}$ and $\mathbf{e}_B=\mathbf{B}/B$  is a unit pseudovector. For propagation parallel to $\mathbf{B}$, the purely electric longitudinal mode\cite{Acosta2016}, with $\mathbf{E}^{(2)}\parallel\mathbf{B}$, yields a nonvanishing pseudoscalar invariant $\mathfrak{G}=\mathbf{E}^{(2)}\cdot\,\mathbf{B}$. This mode lies off the light cone, so that $z_1\equiv k_3^2-\omega^2\neq0$.

The electromagnetic current depends on the invariants $\mathfrak{F}\simeq B^2/2$ and $\mathbf{E}\cdot\mathbf{B}$. For propagation along $\mathbf{B}$, the invariant $\mathbf{E}\cdot\mathbf{B}\neq0$ only for the longitudinal mode $b_\mu^{(2)}$, independent of $C$ symmetry. Expanding $j_\mu$ to linear order in $a_\nu$ yields the electromagnetic current density

\begin{equation}\label{corriente-tensor de conductividad}
j_i = \Pi_{i\nu}a^\nu = Y_{ij}E_j,
\end{equation}

\noindent  where $E_{j}=i(\omega a_{j}-k_{j}a_{0}) $ is the electric field, with $a_{4}=ia_{0}$, $k_{4}=i\omega $, and  $Y_{ij}=\Pi_{ij}/i\omega$ is the complex conductivity. The photon field $a^\mu$ can in general be expanded in the
eigenmodes $b_\mu^{(i)}$. For \(a_\mu=b^{(2)}_\mu\), the eigenvalue equation for \(\Pi_{\mu\nu}\) yields the longitudinal-mode current\cite{Acosta2015,Acosta2016}:

\begin{equation}\label{corriente-escalar s}
j_i = s\,b_i^{(2)},
\end{equation}

\noindent with $s = c^{(2)\mu}\Pi_\mu^\nu c_\nu^{(2)}$ the longitudinal-mode eigenvalue (see (\ref{escalares}) in Appendix \ref{App1}). The scalar $s$ in the one-loop approximation is
\cite{Shabad1,Hugo2,Hugo3,Shabad,Acosta2016}:

\begin{multline}\label{escalar-s}
\begin{split}
  s=&-\frac{e^{3}B}{\pi^{2}}\,\sum_{n=0}^{\infty}\int_{-\infty}^{\infty}\frac{dp_{3}}
  {\varepsilon_{q}}\,[\alpha_{n}\varepsilon^{2}_{n,0}(2p_{3}k_{3}+z_{1})]\\
 &\times\frac{[n^{p}(\varepsilon_{q})+n^{e}(\varepsilon_{q})-1]}{4z_{1}p_{3}(p_{3}
 +k_{3})+z^{2}_{1}-4 \omega^{2} \varepsilon^{2}_{n,0}}\,,
 \end{split}
\end{multline}

\noindent where  $n^{e,p}=[1 + e^{(\varepsilon_{q} \mp \mu)/T}]^{-1}$ are the electron and positron distribution functions in momentum space,  $\varepsilon_{n,0}=\sqrt{m^{2}+|e|Bn}$, with $n=2n_{l}+1-sgn(e) s_{3}$, $\alpha_{n}=2-\delta_{n,0}$ and $q=(n,p_{3})$. Here the $-1$ inside the square brackets accounts for the quantum vacuum limit $(\mu,T)\rightarrow(0,0)$.

Since $b_\mu^{(2)}$ is a pseudovector, the current $j_i$ in (\ref{corriente-escalar s}) is also a pseudovector, a necessary condition for the dynamical generation of a chiral imbalance\cite{Acosta2016}. For the longitudinal mode, fermions exchange momentum $k_3$ with the field while Landau quantum numbers remain unchanged, effectively reducing the problem to a $(1+1)$-dimensional form,
with the lowest Landau level (LLL) providing the strictly one-dimensional limit. In a uniform magnetic field the Dirac equation factorizes for each level $n$. Higher Landau levels can be viewed as copies of the same longitudinal
system with an effective transverse mass $m_{eff} = \sqrt{|e|Bn}$.

A gauge transformation yields $b_3^{(2)}=(k_4/z_1)E_3$ and thus $j_3 = s(k_4/z_1)E_3$ with $E_3=E^{(2)}(\mathbf{e}_B\cdot\mathbf{e}_3)$. Using $\Pi_{33}=s(k_4^2/z_1)$ (from two-dimensional transversality $\Pi_{\mu\nu}k^{\nu}=0$), this becomes\cite{Acosta2016}

\begin{equation}\label{jota3}
j_{3}=\frac{\Pi_{33}}{k_{4}}E_{3}.
\end{equation}

We focus on the real part of $j_3$, since the conductivity is determined by $Im[\Pi_{33}]$  (as will be discussed in Sec. \ref{sec3A}). Using the transversality condition and the two-dimensional identity $\gamma^\mu\gamma^5=-\epsilon^{\mu\nu}\gamma_\nu$ \cite{Peskin1995}, we can write the following relation for the divergence\cite{Acosta2016}  of the axial current $j_A^\mu$

\begin{equation}\label{axial anomaly-1}
  k_\mu j_A^\mu=i\eta\, Im[s]\,E^{(2)},
\end{equation}

\noindent where $\eta=(\textbf{e}_{B}\cdot\textbf{e}_{3})$ is a pseudoscalar factor. Equation (\ref{axial anomaly-1}) has the structure of an axial-current continuity equation with a nonvanishing source term. Thus, the longitudinal electromagnetic mode does not merely probe a chiral response: through its pseudoscalar coupling to $\mathbf{E}^{(2)}\cdot\,\mathbf{B}$, it provides a microscopic mechanism for dynamical axial-charge generation and the associated chiral magnetic current\cite{Acosta2016}. In this sense, the result is structurally analogous to the source term in the Adler--Bell--Jackiw relation\cite{Adler1,Bell}, although here the coefficient is determined by the dissipative longitudinal polarization response of the medium\cite{Acosta2016}.

The one-loop expression\cite{Hugo3} for $Im[s]$  (details in Appendix~\ref{App2}) is:

\begin{align}\label{Imag-s}
Im[s]=-\frac{e^{3}B}{2\pi}\sum_{n=0}^{\infty}\frac{\alpha_{n}\,\varepsilon_{n,0}^{2}}{\Lambda}[\theta_1\,\Delta N(\varepsilon_{r})+\theta_2\,\Delta H(\varepsilon_{s})],
\end{align}

\noindent where $\Lambda=\sqrt{z_{1}(z_{1}+4\varepsilon_{n,0}^{2})}$. Here $\theta_1=\theta(z_1)$ and $\theta_2=\theta(-4\varepsilon_{n,0}^{2}-z_{1})$ are the Heaviside step functions. The term

\begin{align}\label{excitation-particles}
 	\Delta N=[N(\varepsilon_{r})-N(\varepsilon_{r}+\omega)],
 \end{align}

\noindent describes particle excitation ($z_1>0$):$\,[\varepsilon (p_3,n)\longrightarrow \varepsilon (p_3+k_3,n)]$, with $N(\varepsilon_{r})=n^{e}(\varepsilon_{r})+n^{p}(\varepsilon_{r})$. The quantity

\begin{align}\label{creation-particles}
 	\Delta H(\varepsilon_{s})=[H(\varepsilon_{s})+H(\omega-\varepsilon_{s})-2],
 \end{align}

\noindent represents the net pair-creation phase-space factor ($z_{1}<-4\varepsilon^{2}_{n,0}$),  with $H(\varepsilon_{s})=n^{e}(\varepsilon_{s})+n^{p}(\omega-\varepsilon_{s})$; $-2$ is the vacuum limit. The region $-4\varepsilon^{2}_{n,0}<z_{1}<0$ corresponds to a transparency region on the $z_{1}$ axis for real frequencies ($k_{3}^{2}>z_{1}$). The energies $\varepsilon_{s}=(\omega z_{1}+|k_{3}|\Lambda)/2z_{1}$ and
$\varepsilon_{r}=(-\omega z_{1}+|k_{3}|\Lambda)/2z_{1}$ (with $r,s=(n,\omega,k_{3})$) are the
fermion energies in a magnetic field in terms of the longitudinal mode energy $\omega$ and momentum $k_{3}$ (see  (\ref{energia-ecxitacion}) and (\ref{energia-cracion}) in Appendix \ref{App2}).

In the next section, we take the low-energy limit and map the above QED results onto an effective low-energy Dirac model. Within this framework, the longitudinal electromagnetic mode provides the physical motivation for the effective description of field-induced Weyl-like crossing separation and the associated longitudinal magnetotransport response.

\section{EFFECTIVE LOW-ENERGY DIRAC MODEL}
\label{sec3}

We now construct an effective low-energy model for Dirac systems by taking the 
appropriate limit $\omega\to0$, $|k_3|\gg\omega$ from the general QED framework presented in Sec. \ref{sec2}.  Our starting 
point is the Landau spectrum for a three-dimensional Dirac system, characterized 
by the Fermi velocity \(v_{F}\) and an effective Dirac mass \(m^{*}\).  This spectrum is 
obtained from (\ref{espectro/energia}) by the standard replacement \(c \rightarrow v_{F}\) and 
\(m \rightarrow m^{*}\), and in Gaussian CGS units reads

\begin{equation}\label{energia/no campo electrico}
\varepsilon_{n,p_3} = \pm v_F\sqrt{ p_3^2 + m^{*2} v_F^2 + |e|\hbar  B n/c }\,.
\end{equation}

The effective Dirac mass \(m^*\) is retained in the general spectrum and throughout the analytical expressions. It characterizes a possible mass (or gap) scale in the low-energy effective description obtained from the QED framework. Since the present work is not aimed at a specific material, we do not impose a material-dependent value for \(m^*\). For the numerical evaluation, we focus instead on the approximately gapless Dirac regime and take \(m^*\) to be negligibly small on the energy scales relevant to the response. Thus, whereas the numerical results are evaluated in the limit  \(m^*\to0\), the analytical expressions retain  the complete dependence on  $m^*$, so that they remain applicable to nearly gapped Dirac systems. The Weyl-like crossing interpretation adopted below applies to the gapless limit \(m^*\to0\); for finite \(m^*\), the effective spectrum retains the corresponding mass scale.

The longitudinal mode obtained from the polarization operator couples to the magnetized fermion system and induces an orientation-dependent axial response. In the static regime, (\ref{campo electrico pseudovectorial}) gives \(|E_3|/a\simeq |k_3|\), identifying \(\hbar |k_3|\) as the characteristic longitudinal momentum scale associated with the mode.  Motivated by the orientation-dependent axial response obtained above, we introduce an effective low-energy parametrization in which this scale enters as a chirality-dependent longitudinal momentum displacement, \(\Delta=\sigma p_3+\eta\hbar |E_3|/a\), where \(\sigma=\pm1\) labels the two sectors and $\eta=(\textbf{e}_{B}\cdot\textbf{e}_{3}) = \pm 1$ specifies the orientation of the longitudinal electric field relative to the magnetic field. This construction defines the effective spectral model studied below; it does not constitute a microscopic derivation of the fermionic self-energy or establish the displacement directly from the axial-response equation. The resulting spectrum in the effective massless approximation is

\begin{equation}\label{relacion de dispersion en los nodos de Weyl en funcion del momentum de los fermiones}
  \varepsilon_{p_3,n,\sigma}=\pm v_{F}\sqrt{\big(\sigma p_{3}+\frac{\eta\hbar |E_3|}{a}\big)^2+|e|\hbar nB/c},  \hspace{2mm} n\geq1,\nonumber
\end{equation}
\begin{equation}
  \varepsilon_{p_3,0,\sigma}= v_{F}\big(\sigma p_{3}+\frac{\eta\hbar |E_3|}{a}\big), \hspace{2mm}n=0,
\end{equation}

\begin{figure}[h!]
    \centering
    \includegraphics[width=1.0\linewidth]{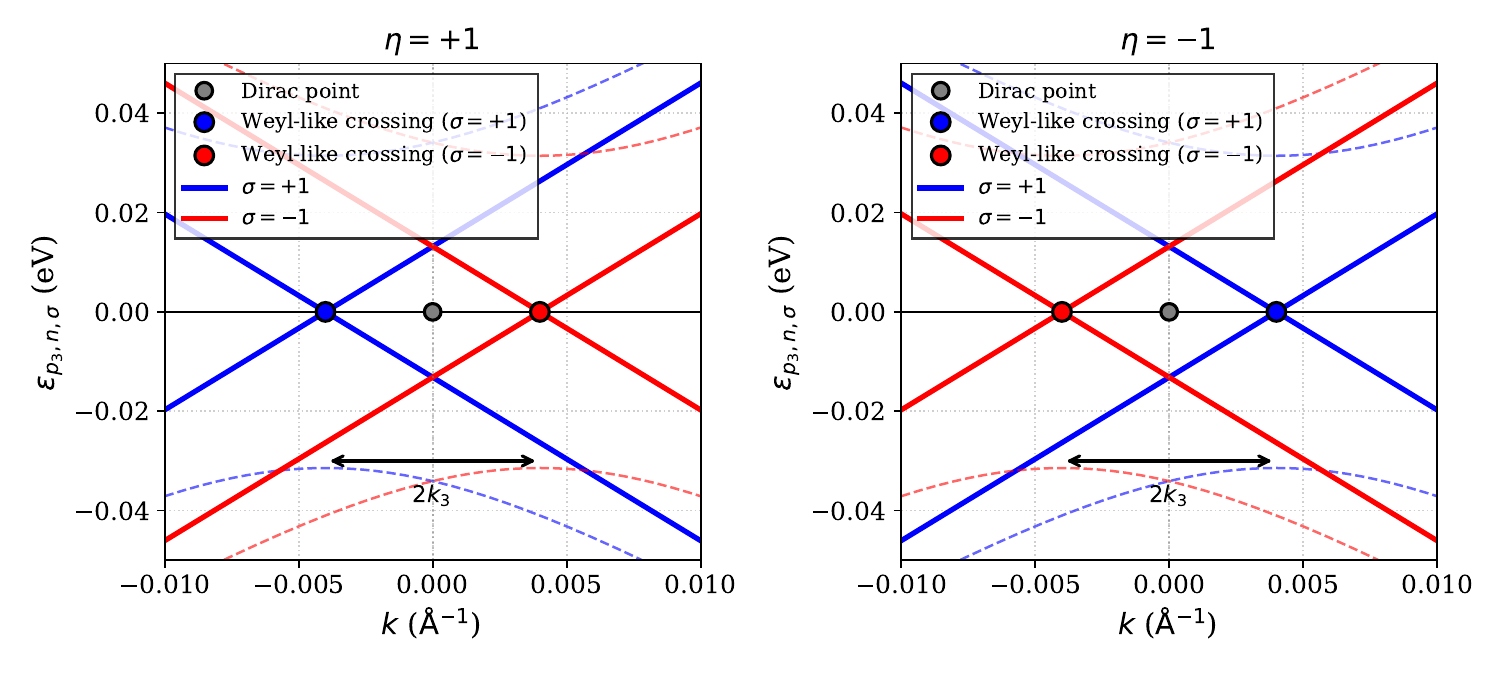}
    \caption\small{{
Low-energy Landau-level spectrum obtained from (\ref{relacion de dispersion en los nodos de Weyl en funcion del momentum de los fermiones}) for the two orientations \(\eta=\pm1\) of the longitudinal electric field. In the massless limit, the coupling to the longitudinal electromagnetic mode produces opposite momentum shifts in the two chiral sectors, separating the zero-energy crossing of the effective Dirac spectrum into two field-induced Weyl-like crossings. These crossings occur at \(k=\pm k_3\), or equivalently at momenta \(p_3=\pm\hbar k_3\). Reversing the orientation, \(\eta\to-\eta\), reverses the direction of the shifts while leaving their separation unchanged. The plotted spectrum illustrates the effective model and is not intended as a material-specific band-structure calculation. The parameters are \(B=6\,\mathrm{T}\),  $v_F=5\times 10^7\, \mathrm{cm/s}$, and \(k_3=0.004\,\text{\AA}^{-1}\).}}
    \label{espectrum}
\end{figure}

 The locations of the gapless crossings follow specifically from the \(n=0\) spectrum,
which vanishes at \(k=\pm k_3\) ($k_3>0$). The levels with \(n\geq1\) remain gapped in this effective spectrum. 
 Thus, the two crossings occur at opposite longitudinal momenta, with a separation \(2k_3\) in momentum space, as illustrated in Fig. \ref{espectrum}.  Within the present effective model, these are identified as Weyl-like crossings associated with the opposite momentum shifts of the two chiral sectors. Establishing their topological character would require an independent analysis of the relevant band structure and its eigenstates, which is beyond the scope of the present work.

These effective crossings are associated with the momentum transfer $k_3$
carried by the longitudinal mode.
Within the QED description, $k_3$ is constrained by the
longitudinal-mode dispersion relation $k_3^2-\omega^2=s(k_3,\omega;B,T,\mu)$, where $s$ (equation (\ref{escalar-s})) is the corresponding eigenvalue of the photon
polarization tensor in the magnetized medium. Thus, solving this dispersion relation for specified thermodynamic
parameters would, in principle, determine the corresponding 
momentum $k_3$ microscopically. In the present low-energy transport analysis, however, we do not solve this dispersion relation explicitly. We therefore do not claim a microscopic numerical determination of $k_3$.

For the numerical illustration, we take
$k_3=0.004\,\text{\AA}^{-1}$, corresponding to a Weyl-like crossing  separation $2k_3=0.008\,\mathrm{\AA}^{-1}$, which is a small fraction of the Brillouin-zone scale in typical Dirac semimetals\cite{Armitage2018}. This value is used as a representative low-energy momentum scale, chosen to illustrate the field-induced splitting within the effective model rather than to describe a specific material. No attempt is made
here to determine $E_3$  from a material-specific
microscopic model. The values $v_F=5\times 10^7\, \mathrm{cm/s}$ and
$B=6\,\mathrm{T}$ are likewise adopted as representative
parameters for illustrating the response of the effective
low-energy model.

It is important to distinguish the field-induced separation of Weyl-like crossings in our effective model from the intrinsic separation of Weyl nodes in equilibrium Dirac materials~\cite{Armitage2018}. In the latter, the node positions are determined by the crystal symmetry and the material-specific band structure. In the present approach, the separation is instead parametrized by the longitudinal momentum scale \(k_3\) associated with the electromagnetic mode. Since \(k_3\) is not determined self-consistently from the mode dispersion in the present calculation, the resulting splitting should be understood as an effective low-energy description of the field-induced displacement of the crossings, rather than as a microscopic prediction of the band-structure modification in a specific material.

\subsection{Longitudinal magnetoconductivity}
\label{sec3A}

Having established the axial response associated with the longitudinal electromagnetic mode through its coupling to the pseudoscalar invariant \(\mathbf{E}^{(2)}\cdot\mathbf{B}\), we now turn to the corresponding transport response. Motivated by this coupling, the effective low-energy model introduced above incorporates opposite longitudinal momentum shifts in the two chiral sectors, giving rise to separated zero-energy crossings in the effective spectrum. Within the present response-function framework, the longitudinal magnetoconductivity is determined by the corresponding eigenvalue \(s\) of the photon polarization tensor in the magnetized medium. We start with the general expression for the real part of the conductivity, which can be written in terms of the imaginary part of  $\Pi_{\mu\nu}$ as \cite{Shabad1,Hugo2,Hugo3,Shabad}:

\begin{equation}\label{conductividad-OP}
  \sigma_{ij}=\frac{Im[\Pi_{ij}]}{\omega}.
\end{equation}

In general, the current density in (\ref{corriente-tensor de conductividad}) can be decomposed as $j_{i}=\sigma_{ij}^{0}E_{j}+(E\times S)_{i}$, where $\sigma_{ij}^{0}=Im[\Pi_{ij}^{S}]/\omega$ and
$S_{i}=\frac{1}{2}\epsilon_{ijk}\sigma_{jk}^{H}$ is a pseudovector associated with the Hall conductivity  $\sigma_{jk}^{H}=Im[\Pi_{ij}^{A}]/\omega$
(where $\epsilon_{ijk}$ is the antisymmetric Levi-Civita tensor). Here $\Pi_{ij}^{S},\Pi_{ij}^{A}$ are the symmetric and antisymmetric
parts of $\Pi_{\mu\nu}$ respectively. 
The first term corresponds to the Ohmic current, while the second describes the Hall current\cite{Lidice}. The Hall and Faraday effects require $C$-nonsymmetric conditions ($\mu \neq 0$). In contrast, the  longitudinal chiral current  persists even in the $C$-symmetric case ($\mu = 0$) \cite{Acosta2016}.  In what follows, we focus exclusively on the Ohmic current contribution.

At this general level, the current contains contributions from both
particle scattering and electron-positron pair creation, which arise from the decay of longitudinal photons and become relevant
at higher frequencies. Since we are interested in the low-energy response of Dirac systems, we take the static limit $\omega \to 0$ with $|k_3| \gg \omega$. In this regime, only particle-hole excitations contribute to the longitudinal current. This limit is consistent with the gauge invariance of $a_{\mu}$ and provides a suitable framework for describing such systems at low energies\cite{Acosta2016}.

Using (\ref{corriente-escalar s}) and applying an appropriate gauge transformation that yields the pseudovector potential $b_{3}^{(2)}=E_3/k_4$, with $E_3=a|k_3|(\textbf{e}_{B}\cdot\textbf{e}_{3})$ (see (\ref{campo electrico pseudovectorial})  in the low-frequency limit), we obtain\cite{Acosta2016}

\begin{equation}\label{corriente longitudinal}
j_3 = \frac{Im[s]}{\omega} E_3, 
\end{equation}

\noindent where $\sigma_B = \sigma_{33}^{0} = Im[s]/\omega$ is the longitudinal magnetoconductivity. Thus, as a result of (\ref{Imag-s}),  and (\ref{corriente longitudinal}), we obtain

\begin{equation}\label{corriente-limite-low energy}
\begin{split}
  j_{3}=&
  \frac{e^{3}|k_{3}|}{4\pi z_{1}}\frac{\sinh(\frac{\omega}{2T})}{\omega}
   \sum_{n=0}^{n_{\mu,T}}\bigg[\frac{\alpha_n\varepsilon_{n,0}^2}{\lambda}
   \frac{1}{\cosh(\frac{\omega}{2T})+
   \cosh(\frac{\lambda-\mu}{T})}\\&+\frac{1}
   {\cosh(\frac{\omega}{2T})+\cosh(\frac{\lambda+\mu}{T})}\bigg](\textbf{E}^{(2)}\cdot \textbf{B}),
   \end{split}
\end{equation}

\noindent where 

\begin{equation}
\lambda=\frac{|k_{3}|}{2}\sqrt{1+\frac{4\varepsilon_{n,0}^2}{z_{1}}},
\end{equation}

\noindent and $\textbf{E}^{(2)}=E^{(2)}\textbf{e}_{3}$. We assume $\Delta_B\sim \mu$ and $\Delta_B \gg T$, so that higher Landau levels can be neglected, where $n_{\mu,T}$ denotes the contribution from the lowest Landau
levels under the conditions stated above. The second term in (\ref{corriente-limite-low energy}) corresponds to
hole contributions from positively charged carriers, arising from the $\varepsilon_q + \mu$ term in the Fermi-Dirac distribution. Taking the static limit $\omega \to 0$, (\ref{corriente-limite-low energy}) reduces to\cite{Acosta2016}:

\begin{equation}\label{corriente-limite-low energy-omega cero}
\begin{split}
  j_{3}=&
  \frac{e^{3}}{8\pi|k_{3}|T}
   \sum_{n=0}^{n_{\mu,T}}\bigg[\frac{\alpha_n\varepsilon_{n,0}^2}{\lambda}\frac{1}{1+
   \cosh(\frac{\lambda-\mu}{T})}\\&+\frac{1}
   {1+\cosh(\frac{\lambda+\mu}{T})}\bigg](\textbf{E}^{(2)}\cdot \textbf{B}),
   \end{split}
\end{equation}

\noindent where $\textbf{E}^{(2)}=a|k_3|\textbf{e}_{3}$. Notably,  (\ref{corriente-limite-low energy-omega cero}) remains finite as $\mu \to 0$, indicating that the chiral current persists even in the $C$-symmetric case\cite{Acosta2016}. This distinguishes the longitudinal chiral current from the Faraday rotation and Hall current, which vanish when $\mu = 0$\cite{Lidice}.

From (\ref{corriente longitudinal}) and the expression (\ref{corriente-limite-low energy-omega cero}) (restoring units of $c$ and $\hbar$) the longitudinal magnetoconductivity at finite temperature $T$ and chemical potential $\mu$ takes the form

\begin{equation}\label{magnetoconductividad}
\begin{split}
  \sigma_B=&\frac{e^3 B v_F^3  m^{*2} }{4\pi\hbar^2 k_3 T}\bigg[\frac{(2\varepsilon_0)^{-1}}{1+\cosh[\frac{(\varepsilon_0-\mu)}{T}]}
  +\frac{\varepsilon_{1}^{-1}}{1+\cosh[\frac{(\varepsilon_1-\mu)}{T}]}\bigg]\\
  &+\frac{e^4 B^2 v_{F}^2}{2\pi\hbar c^2 k_3 T}\bigg[\frac{\varepsilon_{1}^{-1}}{1+\cosh[\frac{(\varepsilon_1-\mu)}{T}]}\bigg],
\end{split}
\end{equation}

\noindent where
\begin{equation}\label{energia-estado basico}
  \varepsilon_0=\frac{1}{2}v_F\sqrt{\hbar^2k_{3}^{2}+4m^{*2}v_F^2}\;,
\end{equation}
\begin{equation}\label{energia-n=1}
  \varepsilon_1=\frac{1}{2}v_F\sqrt{\hbar^2k_{3}^{2}+4m^{*2}v_F^2+8eB\hbar/c}\;.
\end{equation}

Here, we retain only the electronic sector and truncate the sum over Landau levels at \(n=1\), since the thermal population of higher levels is strongly suppressed in the parameter regime under consideration. The occupation of the \(n=2\) level remains below \(1.01\%\) for \(T\leq 30\,\mathrm{K}\) and \(B>4\,\mathrm{T}\), consistent with the hierarchy between the thermal energy and the characteristic magnetic energy scale. This provides a physical motivation for the low-Landau-level approximation adopted in evaluating the longitudinal conductivity. Equation (\ref{magnetoconductividad})  provides a unified expression for the longitudinal
magnetoconductivity, with the conventional and chiral contributions
arising within the same electromagnetic-response framework. The present response-function calculation does not introduce an independent phenomenological internode relaxation time. It therefore provides a complementary description of the longitudinal response, rather than a complete account of disorder-dependent steady-state transport.

The terms $\varepsilon_0$ and $\varepsilon_1$ denote effective particle-hole excitation thresholds associated with the $n=0$ and $n=1$ Landau levels, rather than the corresponding single-particle dispersions. They depend on $m^*$, $B$, and the field-induced momentum scale $k_3$. In our approach this momentum acts as a field-dependent parameter separating the zero-energy crossing of the effective Dirac spectrum into two field-induced Weyl-like crossings and modulates the Fermi surface via $[1+\cosh((\varepsilon_{0,1}-\mu)/T)]^{-1}$.

We emphasize that the limits \(\omega\to0\) and \(k_3\to0\) do not commute in the present longitudinal response. Accordingly, we define the conductivity considered here as the finite-wave-vector longitudinal response evaluated in the regime \(|k_3|\gg\omega\), where the longitudinal mode lies far from the light cone. The finite value of \(k_3\) is set by the longitudinal  mode discussed above and is therefore retained when taking the low-frequency limit. Equation (\ref{magnetoconductividad}) should consequently be understood as a finite-wave-vector longitudinal magnetoconductivity, rather than as the homogeneous \(k_3\to0\) DC conductivity. No extrapolation to \(k_3\to0\) is implied.

\begin{figure}
    \centering
    \includegraphics[width=1.0\linewidth]{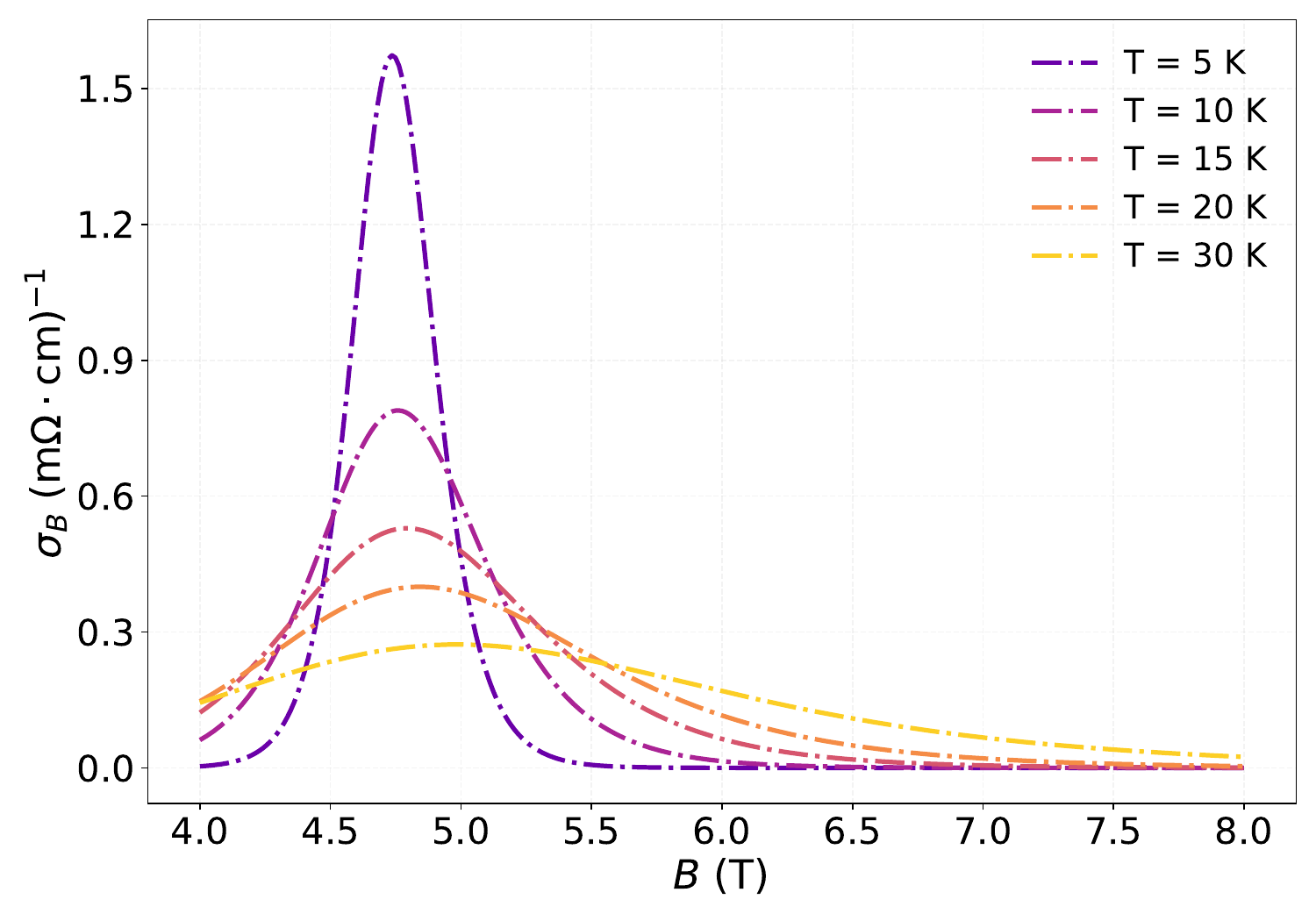}
    \caption\small{{
    Longitudinal magnetoconductivity $\sigma_B$ as a function of the applied magnetic field $B$ for several temperatures in the massless limit. The parameters are $v_F=5\times 10^7\, \mathrm{cm/s}$, $k_3=0.004\,\text{\AA}^{-1}$, and $\mu=40\,\mathrm{meV}$. The nonmonotonic field dependence originates from the variation of the Landau-level threshold $\varepsilon_1(B)$  with $B$, which enters the occupation factors. For the parameters considered, $\varepsilon_1(B)$ crosses the chemical potential near $B\simeq4.7\,\mathrm{T}$, giving rise to the pronounced low-temperature feature. 
    }}
    \label{conduct/B}
\end{figure}

Separating the conventional and chiral terms in
(\ref{magnetoconductividad}), we write the
longitudinal magnetoconductivity as
\begin{equation}
\label{conductividad-comportamientos}
\sigma_B(B,T)
=
\sigma_{\rm conv}(B,T)
+
B^2 C_{\rm eff}(B,T),
\end{equation}
where $C_{\rm eff}(B,T)$ is the effective coefficient
of the chiral contribution. Unlike the semiclassical
Son--Spivak description~\cite{Spivak2013}, in which
the anomalous conductivity depends on an internode
relaxation time, the present response-function
calculation does not introduce an independent
phenomenological relaxation time. Our QED-based response-function framework determines this contribution from the same electromagnetic response that generates the field-induced momentum shift \(k_3\), while retaining the Landau-level structure and finite-mass corrections. The coefficient $C_{\rm eff}(B,T)$ depends on the effective momentum
scale $k_3$, the Landau-level spectrum, and the
corresponding Fermi--Dirac occupation factors.
Because this coefficient itself depends on $B$,
(\ref{conductividad-comportamientos}) does not
imply a global quadratic dependence of the
magnetoconductivity on the magnetic field.

\begin{figure}
    \centering
    \includegraphics[width=1.0\linewidth]{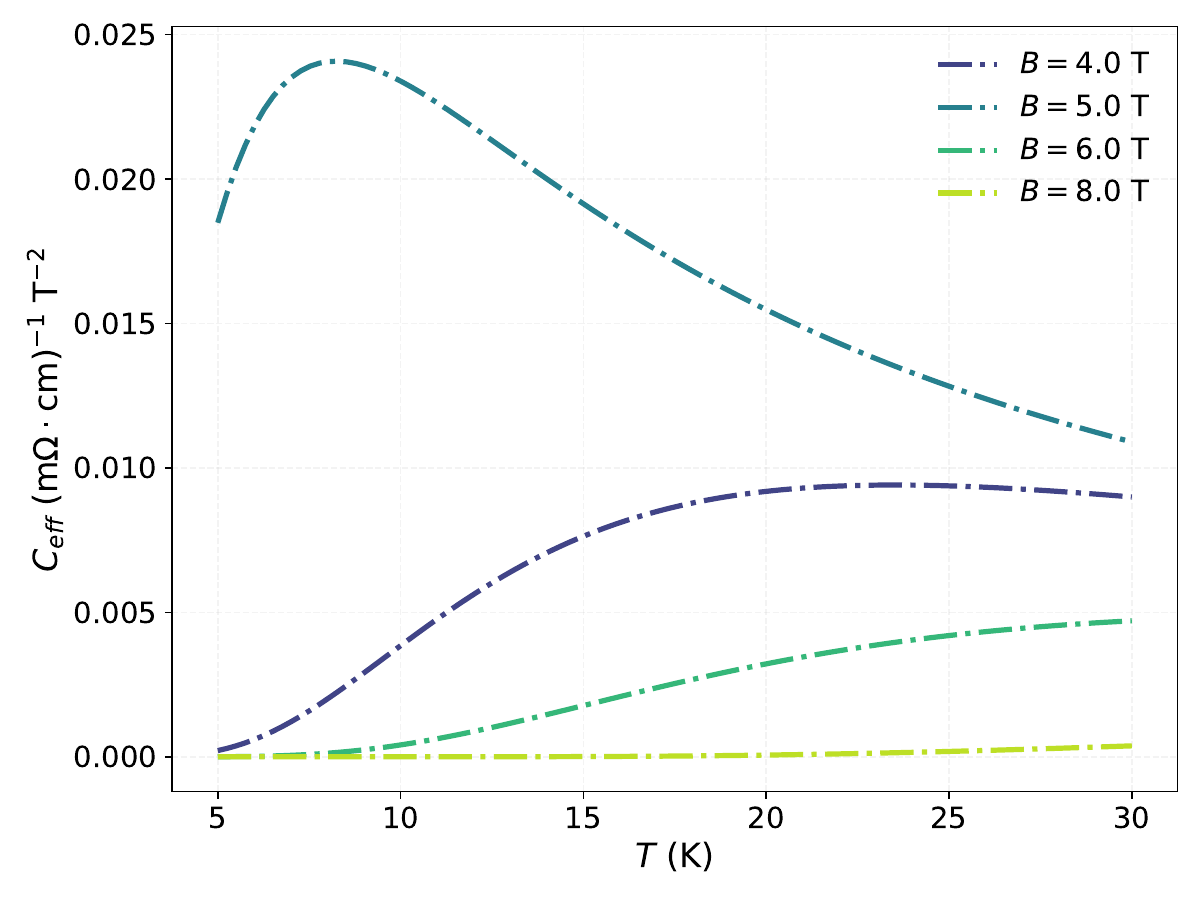}
    \caption\small{{Effective coefficient $C_{\mathrm{eff}}(B,T)=\sigma_{\mathrm{ch}}(B,T)/B^2$ of the chiral contribution to the longitudinal magnetoconductivity as a function of temperature for several magnetic fields. The field dependence of $C_{\mathrm{eff}}$ reflects how the Landau-level threshold $\varepsilon_1(B)$ and the associated Fermi–Dirac occupation factors vary with $B$. Thus, the $B^2$ dependence represents an explicit prefactor of the chiral contribution rather than a global quadratic scaling law for the full magnetoconductivity. The parameters are $v_F=5\times 10^7\, \mathrm{cm/s}$, $k_3=0.004\,\text{\AA}^{-1}$, and $\mu=40\,\mathrm{meV}$.
}}
    \label{fig:placeholder}
\end{figure}

Fig.~\ref{conduct/B} shows $\sigma_B$ as a
function of magnetic field for several temperatures
in the massless limit. The nonmonotonic
low-temperature response is governed by the
field-dependent excitation threshold

\begin{equation}
\label{energia-B}
\varepsilon_1(B)
=
\frac{v_F}{2}
\sqrt{
\hbar^2 k_3^2
+
\frac{8|e|\hbar B}{c}
}.
\end{equation}
For the chosen parameters, this threshold crosses
the chemical potential near $B\simeq 4.7~\mathrm{T}$,
giving rise to the pronounced feature in
Fig.~\ref{conduct/B}. The corresponding coefficient,
$C_{\rm eff}(B,T)=\sigma_{\rm ch}(B,T)/B^2$,
is plotted in Fig.~\ref{fig:placeholder}.
Its magnetic-field dependence reflects the
evolution of $\varepsilon_1(B)$ and the associated
occupation factors.

An inverse-temperature dependence follows from
(\ref{magnetoconductividad}) within the
electron-sector approximation, provided that
\begin{equation}
|\varepsilon_1-\mu|
\ll T
\ll \varepsilon_2-\mu,
\qquad
\frac{\varepsilon_1+\mu}{T}\gg 1.
\end{equation}
Here 
$\varepsilon_2$ denotes the excitation threshold
of the next neglected Landau level. The first
condition allows the occupation factor of the
$n=1$ level to be expanded, while the remaining
conditions suppress higher-level and hole
contributions. In this intermediate asymptotic regime,
we obtain

\begin{equation}
\label{eq:Ceff_highT}
C_{\mathrm{eff}}(T,B)
\simeq
\frac{e^4 v_F^2}
{4\pi\hbar c^2 k_3\varepsilon_1 T}.
\end{equation}
The $1/T$ dependence originates from the thermal
broadening of the Fermi--Dirac distribution.
More generally, (\ref{corriente-limite-low energy-omega cero})
also yields inverse-temperature scaling when
$T\gg|\lambda-\mu|$ and
$T\gg|\lambda+\mu|$, so that both electron
and hole occupation factors approach constants.
This formal limit must be distinguished from
the restricted temperature window in which
(\ref{eq:Ceff_highT}) and the few-Landau-level
approximation apply.

The temperature dependence of $C_{\rm eff}(B,T)$  is
shown in Fig.~\ref{fig:placeholder}. For fixed
$B$ and $k_3$, its electronic contribution
satisfies
\begin{equation}
C_{\rm eff}(B,T)
\propto
\frac{1}{
T\left[
1+\cosh\left(
\frac{\varepsilon_1(B)-\mu}{T}
\right)
\right]
}.
\end{equation}
For $\varepsilon_1(B)\neq\mu$, this function
has a maximum at
$T_M\simeq 0.65|\varepsilon_1(B)-\mu|$,
giving rise to a relatively weak temperature
dependence near $T_M$. The position of this
maximum changes with $B$, $\mu$, and $k_3$.
Consequently, the correlated magnetic-field
and temperature dependence of $C_{\rm eff}(B,T)$,
particularly near the Landau-level threshold,
provides a characteristic prediction of
the effective model.

\section{CONCLUSIONS}
\label{sec4}

Our analysis establishes a connection, within an effective low-energy description, between the longitudinal electromagnetic response of a magnetized Dirac system and the field-induced separation of Weyl-like crossings. Starting from the one-loop photon polarization tensor in a magnetized fermionic medium, we identify a longitudinal electromagnetic response associated with a nonvanishing pseudoscalar invariant \(\mathbf{E}^{(2)}\cdot\mathbf{B}\). Motivated by this coupling, we formulate an effective low-energy description in which an axial coupling produces opposite momentum shifts in the two chiral sectors. In the massless limit, these shifts separate the zero-energy crossings of the  spectrum, giving rise to an effective chiral band structure with Weyl-like crossings. Their topological nature, however, is not established in the present work.

The same response-function framework yields
an analytical expression for the longitudinal
magnetoconductivity at finite temperature
and chemical potential, without introducing
an independent phenomenological internode
relaxation time. It therefore provides a complementary description of the longitudinal response, rather than a complete account of disorder-dependent steady-state transport. The resulting conductivity
contains conventional and chiral contributions, with
the latter exhibiting an explicit $B^2$
prefactor with a field- and temperature-dependent
coefficient $C_{\mathrm{eff}}(B,T)$.
Landau-level quantization and the associated
occupation factors produce a nonmonotonic
magnetic-field dependence, particularly
near the threshold at which the first
excited Landau level crosses the chemical
potential. Consequently, the full
magnetoconductivity does not generally
follow a quadratic magnetic-field law.

The temperature dependence of the chiral
contribution provides an additional
characteristic of the proposed mechanism.
The effective coefficient
$C_{\mathrm{eff}}(B,T)$ exhibits a broad
maximum associated with the thermal
occupation of the first excited Landau
level, giving rise to an approximately
temperature-independent region.
At sufficiently high temperatures within
the regime of validity of the model,
the coefficient approaches an inverse-
temperature dependence. The position
and magnitude of these features depend
on the magnetic field, chemical potential,
and effective momentum scale $k_3$,
providing correlated field- and
temperature-dependent signatures of
the longitudinal response.

The present results describe the
longitudinal response of an effective
Landau-quantized Dirac model rather
than the complete transport behavior
of a specific material. A quantitative
comparison with experiments will require
a self-consistent determination of the
longitudinal-mode momentum scale,
together with material-specific band
structures and the effects of disorder
and relaxation. Extending the present
framework in these directions would
help establish the experimental conditions
under which field-induced chiral
band structure and its associated
magnetotransport signatures can
be identified.


\appendix

\section{}

\subsection{Electromagnetic modes for propagation along \textbf{B}}
\label{App1}

The structure of $\Pi_{\mu\nu}$ in a magnetized relativistic quantum medium allows us to determine the polarization properties of three electromagnetic eigenmodes propagating in the system \cite{Hugo2,Hugo3,Shabad1}. Under these conditions, the polarization tensor can be expanded in
 terms of six independent transverse tensors \cite{Hugo3}

 \begin{equation}\label{diagonalize form of the polarization tensor}
   \Pi_{\mu\nu}=\overset{6}{\underset{n=1}{\sum}}\pi^{(i)}\Psi_{\mu\nu}^{(i)}.
 \end{equation}

As shown in Ref.\cite{Hugo2}, symmetry properties of the quantum-statistical
system reduce the number of independent basic tensors from nine to six. The basic tensors are:

\begin{eqnarray}\label{tens-basicos-S}
  \Psi_{\mu\nu}^{(1)} &=& k^{2}T_{\mu\nu},\quad\Psi_{\mu\nu}^{(2)}=(Fk)_{\mu}(Fk)_{\nu}\nonumber\\
  \Psi_{\mu\nu}^{(3)} &=& -k^{2}T_{\mu}^{\lambda} F_{\lambda\eta}^{2}T^{\eta}_{\nu},\quad\Psi_{\mu\nu}^{(4)}=R_{\mu}R_{\nu},
\end{eqnarray}

\noindent with $T_{\mu\nu}=(g_{\mu\nu}-k_{\mu}k_{\nu}/k^{2})$ and $R_{\mu}=(u_{\mu}-k_{\mu}(uk)/k^{2})$, where $u_{\mu}$ is the four-velocity of the medium and $F_{\mu\nu}$ is the electromagnetic tensor. We express tensor quantities in Minkowski metric, in the form $g_{\mu\nu}=(1,1,1,-1)$, where the index ordering $\mu,\nu=1,2,3,0$ is used, consistent with
the convention adopted in Sec. \ref{sec2} . The tensors in (\ref{tens-basicos-S}) are symmetric in $\mu$ and $\nu$,  while the following ones are antisymmetric

\begin{eqnarray}
\Psi_{\mu\nu}^{(5)}&=&(uk)[k_{\mu}(Fk)_{\nu}-k_{\nu}(Fk)_{\mu}+k^{2}F_{\mu\nu}]\nonumber\\
\Psi_{\mu\nu}^{(6)}&=&u_{\mu}(Fk)_{\nu}-u_{\nu}(Fk)_{\mu}+(uk)F_{\mu\nu}.
\end{eqnarray}\label{tens-basicos-A}

We introduce a set of orthonormal vectors which are the
  eigenvectors of $\Pi_{\mu\nu}$ in the limit $(\mu,T)\rightarrow(0,0)$

\begin{eqnarray}\label{vect-orton}
c_{\mu}^{(1)}&=& R_{1}(F^{2}k)_{\mu}k^{2}-k_{\mu}(kF^{2}k),\quad c_{\mu}^{(2)}=R_{2}(F^{*}k)_{\mu}\nonumber\\
c_{\mu}^{(3)}&=& R_{3}(Fk)_{\mu},\quad c_{\mu}^{(4)}= R_{4}k_{\mu},
\end{eqnarray}

\noindent where $R_{i}\, (i=1,2,3,4)$ are normalization parameters, and $F^{*}_{\mu\nu}$ is the dual of the electromagnetic tensor
$F_{\mu\nu}$. Using these vectors we can obtain the scalars:

\begin{eqnarray}\label{escalares}
  p&=& c^{(1)\mu}\Pi_{\mu}^{\nu}c^{(1)}_{\nu},\quad s=c^{(2)\mu}\Pi_{\mu}^{\nu}c^{(2)}_{\nu}\\
  t&=& c^{(3)\mu}\Pi_{\mu}^{\nu}c^{(3)}_{\nu},\quad r=c^{(3)\mu}\Pi_{\mu}^{\nu}c^{(1)}_{\nu}
\end{eqnarray}

and the pseudoscalars:

\begin{subequations}\label{psudoescalares}
\begin{eqnarray}
  q&=&c^{(2)\mu}\Pi_{\mu}^{\nu}c^{(1)}_{\nu},\quad v=c^{(2)\mu}\Pi_{\mu}^{\nu}c^{(3)}_{\nu}.
\end{eqnarray}
\end{subequations}

From (\ref{diagonalize form of the polarization tensor}) we can find the polarization properties\cite{Hugo2,Shabad1} of three electromagnetic eigenmodes of $\Pi_{\mu\nu}$. For propagation along \textbf{B}, the eigenmodes of $\Pi_{\mu\nu}$ are:

\begin{eqnarray}\label{eigenmodes}
b^{(2)}_{\mu}&=&ac^{(2)}_{\mu}\hspace{1mm}e^{i(k_{3}x_{3}-\omega t)}\nonumber\\
b^{(1,3)}_{\mu}&=&b(c^{(1)}_{\mu}\pm ic^{(3)}_{\mu})\hspace{1mm}e^{i(\textbf{k}_{\bot}\cdot\textbf{r}_{\bot}-\omega t)},
\end{eqnarray}

\noindent with eigenvalues $\eta^{(2)}=s$ and $\eta^{(1,3)}=t\pm \sqrt{-r^2}$ respectively. Here $\textbf{r}_{\bot}$ is the coordinate vector in the plane-$(x,y)$, and $a,b$ are parameters, which have dimensions of vector potential. The electric and magnetic fields associated with these modes are obtained from the equations:

\begin{equation}\label{ecuaciones-campo electrico-magnetico}
  \textbf{E}^{(i)}=-\frac{\partial\textbf{b}^{(i)}}{\partial
x_{0}}-\frac{\partial\ b^{(i)}_{0}}{\partial\textbf{x}},\hspace{3mm} \textbf{H}^{(i)}=\nabla\times \textbf{b}^{(i)},
\end{equation}

\noindent with $(i=1,2,3)$. In the 
$C$-symmetric $(\mu=0)$ case, the mode $b_{\mu}^{(3)}$ is a transversely polarized plane wave, whose electric unit vector is
$\textbf{E}_{u}^{(3)}=\textbf{e}_{\perp}\times \textbf{e}_{3}$ ,
orthogonal to the plane $(\textbf{B}, \textbf{k})$, where we define $\textbf{e}_{\perp}=\textbf{k}_{\perp}/k_{\bot}$ and
$\textbf{e}_{3}=\textbf{k}_{3}/k_{3}$ as the transverse and
parallel unit vectors respectively. The mode $b_{\mu}^{(2)}$ is pure electric and
longitudinal with $\textbf{E}_{B}=a (k_3^2-\omega^2)^{1/2}\mathbf{e}_B$ (we recall that $\textbf{e}_{B}=\textbf{B}/B$ is a unit pseudovector), whereas
$b_{\mu}^{(1)}$ is transverse $E_{u}^{(1)}=\textbf{e}_{\perp}$.  In this $C$-symmetric case  $\eta^{(1)}=\eta^{(3)}$ ,
and the circular polarization unit vectors $(\textbf{E}_{u}^{(1)}\pm
i\textbf{E}_{u}^{(3)})/\sqrt{2}$ are common eigenvectors of $\Pi_{ij}$
and of the rotation generator matrix $A^{3ij}$ .

For a $C$-nonsymmetric medium $(\mu\neq 0)$. The second mode
$b_{\mu}^{(2)}$ is the same pure longitudinal wave as in the $C$-symmetric case. The
transverse modes $b_{\mu}^{(1,3)}$ describe circularly polarized
waves in the plane orthogonal to $\textbf{B}$  having different
eigenvalues, typical of the Faraday effect\cite{Lidice}.

\subsection{Calculation of $Im[s]$}
\label{App2}

The denominator $D$ appearing in the integral for $s$ in (\ref{escalar-s}),
which contains the relevant singularities, is given by

\begin{equation}\label{Denominador-D}
  D=4z_{1}p_{3}(p_{3}+k_{3})+z^{2}_{1}-4 \omega^{2}\varepsilon^{2}_{n,0},
\end{equation}

\noindent where $z_{1}=k_{3}^{2}-\omega^{2}$ and $\varepsilon_{n,0}^{2}=m^{2}+|e|nB$, with $n=2n_{l}+1-sgn(e) s_{3}$. It can be written in a form symmetric under the exchange\cite{Hugo3} $\varepsilon_{q}\leftrightarrow \varepsilon_{q^{\prime}}$, $\omega\leftrightarrow-\omega$ 

\begin{multline}\label{denomonador-DI}
 D^{-1}= \frac{1}{8\varepsilon_{q^{\prime}}\varepsilon_{q}\omega} \,(\frac{1}{\varepsilon_{q^{\prime}}-\varepsilon_{q}-\omega+i\epsilon}
  -\frac{1}{\varepsilon_{q^{\prime}}-\varepsilon_{q}+\omega+i\epsilon} \\
 -\frac{1}{\varepsilon_{q^{\prime}}+\varepsilon_{q}-\omega+i\epsilon}+
\frac{1}{\varepsilon_{q^{\prime}}+\varepsilon_{q}+\omega+i\epsilon}),
\end{multline}

\noindent where $\varepsilon_{q^{\prime}}=\sqrt{(p_{3}+k_{3})^2+m^2+|e|nB}$ and $\varepsilon_{q}=\sqrt{p_{3}^{2}+m^2+|e|nB}$, with $q=(n,p_{3})$. The first pair of singularities  are related to excitation of particles to higher energies and the second two are connected to the pair creation. We introduce an infinitesimal positive imaginary part $i\epsilon$
in $\omega$, and by using the relation:

\begin{equation}\label{delta}
   \frac{1}{s-\omega-i\epsilon}=P\frac{1}{s-\omega}+i\pi\delta(s-\omega),
\end{equation}

\noindent where $P$ corresponds to the principal value in the expression, we get for the imaginary  part\cite{Hugo3} of  $D^{-1}$ 

\begin{multline}\label{ImD}
  Im D^{-1}=\pm\, \frac{\pi}{8\varepsilon_q\varepsilon_{q^{\prime}}\omega}\,[\delta(\varepsilon_{q^{\prime}}-\varepsilon_q\mp \omega)+\delta(\varepsilon_{q^{\prime}}-\varepsilon_q\pm \omega) \\
 -\delta(\varepsilon_{q^{\prime}}+\varepsilon_q\mp \omega)],
\end{multline}

\noindent where the $\pm$ signs apply respectively to $\omega\gtrless0$. We can use now  (\ref{ImD}) to obtain the imaginary part of the scalar $s$ (see (\ref{escalar-s}))
according to the relation:

\begin{equation}
\int_{-\infty}^{\infty}dp_3f(p_3)\delta(g(p_3))=\sum_m\frac{f(p_3^m)}{\mid g^{\prime}(p_3^m)\mid}\label{formula},
\end{equation}

\noindent where $p_3^m$, with $m=(1,2)$ are the roots of $g(p_3)=0$. For $p_3 \in\, (-\infty,\infty)$, the denominator in (\ref{escalar-s}) can vanish\cite{Hugo3} only for real $z_{1}$. Thus, the integral in (\ref{escalar-s}) represents an analytic function in the $z_{1}$ plane except possible singularities located  on the real axis, corresponding to the absorption region ( $Im[\Pi_{33}]$ is responsible for the absorption process for the longitudinal mode), where

\begin{equation}\label{momentum}
  p_{3}^{(1,2)}=\frac{-k_{3}z_{1}\pm \omega\Lambda}{2z_{1}},
\end{equation}

\noindent are the roots\cite{Hugo3} of the denominator in (\ref{escalar-s})  and $\Lambda=\sqrt{z_{1}(z_{1}+4\varepsilon^{2}_{n,0})}$. For the present case, $g(p_3)=\omega\pm(\varepsilon_{q^{\prime}}\pm \varepsilon_{q})$, so that

\begin{equation}\label{balance energia/momentum}
  \omega=\varepsilon_{q^{\prime}}\pm \varepsilon_{q},\hspace{2mm}k_{3}=p^{\prime}_{3}\pm p_{3},
\end{equation}

\noindent and the corresponding values of the energies are given by:

\begin{equation}\label{energia-ecxitacion}
  \varepsilon_{r}=\frac{-\omega z_{1}+|k_{3}|\Lambda}{2z_{1}},
\end{equation}
\begin{equation}\label{energia-cracion}
  \varepsilon_{s}=\frac{\omega z_{1}+|k_{3}|\Lambda}{2z_{1}},
\end{equation}

\noindent where $r,s=(n,\omega,k_{3})$. The $\pm$ signs in (\ref{balance energia/momentum}) correspond to the pair creation $(\varepsilon_{s})$ and excitation cases $( \varepsilon_{r})$ respectively. Substitution then gives

\begin{equation}
\mid \frac{d}{dp_3}(g(p_3))\mid=\frac{\Lambda}{2\varepsilon_q^m\varepsilon_{q^{\prime}}^m}
\label{resultado}.
\end{equation}

In evaluating the integral in (\ref{escalar-s}) involving the second delta function in (\ref{ImD}), we make the substitutions $p_{3}+k_{3}\leftrightarrow -p_{3}$, $n^{\prime}\leftrightarrow n$.

\end{document}